\documentclass[preprint,amsmath,amssymb,pre,superscriptaddress]{revtex4-1}
\usepackage{color,longtable}
\usepackage{graphicx}
\usepackage{amssymb,amsfonts,amsmath}
\usepackage{here}

\begin{document}
\title{Spontaneous formation of Frenkel defects in high-entropy-alloys-type compound}
\author{Rikuya Ishikawa}
\affiliation{%
Department of Physics, Tokyo Metropolitan University, 1-1 Minamioosawa, Hachiouji-shi, Tokyo 192-0397, Japan
}
\author{Kyohei Takae}
\affiliation{%
Department of Fundamental Engineering, Institute of Industrial Science, University of Tokyo, 4-6-1 Komaba, Meguro-ku, Tokyo 153-8505, Japan
}%
\author{Yoshikazu Mizuguchi}
\affiliation{%
Department of Physics, Tokyo Metropolitan University, 1-1 Minamioosawa, Hachiouji-shi, Tokyo 192-0397, Japan
}
\author{Rei Kurita}
\affiliation{%
Department of Physics, Tokyo Metropolitan University, 1-1 Minamioosawa, Hachiouji-shi, Tokyo 192-0397, Japan
}%
\date{\today}
\begin{abstract}
High-entropy alloys (HEAs) are attracting attention due to their exceptional properties, such as enhanced mechanical toughness, superconducting robustness, and thermoelectric performance. 
Numerous HEAs have been developed for diverse applications, ranging from self-healing in fusion reactors to addressing environmental concerns with thermoelectric materials. 
Understanding atomic diffusion within HEA crystals is crucial for these applications. 
Here, this study investigates diffusion mechanisms in PbTe-based HEAs, focusing on the role of indium (In). Molecular dynamics simulations reveal that In inclusion prompts spontaneous Frenkel defect formation, notably enhancing diffusion not only of In$^+$ but also other cations. 
Frenkel defect formation, closely linked to alloy properties, is predominantly influenced by charge rather than cation size. 
This insight not only enhances comprehension of HEA diffusion mechanisms but also develops HEAs with properties such as self-healing from damage and high ion permeability, advancing the field of material science.
\end{abstract}


\maketitle

Crystals that incorporate more than five distinct elements, known as high-entropy alloys (HEAs)~\cite{Senkov2017} , have emerged as a significant focus of development for their application in high-temperature structural materials~\cite{Senkov2011, Senkov2010}, medical implants~\cite{Todai2017}, and catalysts~\cite{Kusada2020, Kusada2022}. Characterized by their exceptional superiority over conventional substances, HEAs exhibit remarkable properties~\cite{Yeh2004, Cantor2004, Senkov2017, George2019}. For instance, the Cantor alloy, a combination of Cr, Mn, Fe, Co, and Ni, has the higher fracture toughness among metallic materials~\cite{Gludovatz2014}. In the context of fusion reactors, the superconducting magnets create powerful magnetic fields that confine ultra-high temperature plasma, facilitating fusion reactions. However, this process produces particle beams that can damage the crystal structure of the superconducting magnets. HEAs offer a unique advantage in this scenario due to their self-healing capabilities from such damage~\cite{Yoshiie2000, Zinkle2013}. Additionally, the nonlinear effects known as cocktail effects contribute to significantly enhanced catalytic performance in HEA nanoparticles~\cite{Kusada2020, Kusada2022}. PbTe-based multicomponent materials, in particular, exhibit high thermoelectric performance, making them promising solutions for energy-related challenges~\cite{Jiang2021, Jiang2021-2, Jiang2022, Yamashita2021}. The thermoelectric performance index ($ZT$) of PbTe-based materials has been shown to increase to 2.0 at 900 K~\cite{Jiang2021}, underscoring the exceptional capabilities of HEAs for industrial applications.

Recently, materials with high lithium-ion permeability and melting temperatures are desirable for use as separators in lithium-ion batteries to improve their safety. Consequently, developing HEAs with high diffusion coefficients for specific ions is of great interest.
However, the vast number of potential combinations of components, mixing ratios, and atomic types for creating high-entropy compounds presents a complex challenge. 
The macroscopic strength and crystal stability in HEAs are closely linked to  the diffusion of atoms~\cite{Tsai2008, Tsai2011}. 
Thus, elucidating the mechanisms that enhance diffusion in HEAs is crucial for providing guidelines for material design. 
Previous studies employ molecular dynamics simulations with modified-EAM potentials~\cite{Daw1993}, machine-learning potentials~\cite{Liu2023}, and first-principles calculations~\cite{Feng2017} to investigate diffusion~\cite{Wang2022}, stability~\cite{Widom2014}, and deformation~\cite{Li2016}  in HEA systems, specifically focusing on Cantor alloys. Although static properties have been explained, discrepancies exist between simulation results and experimental observations regarding diffusion~\cite{Senkov2017, Dabrowa2019, Wang2022}, highlighting the need for further investigation into the diffusion mechanisms of multicomponent crystals.

\begin{figure}[htbp]
\centering
\includegraphics[width=8cm]{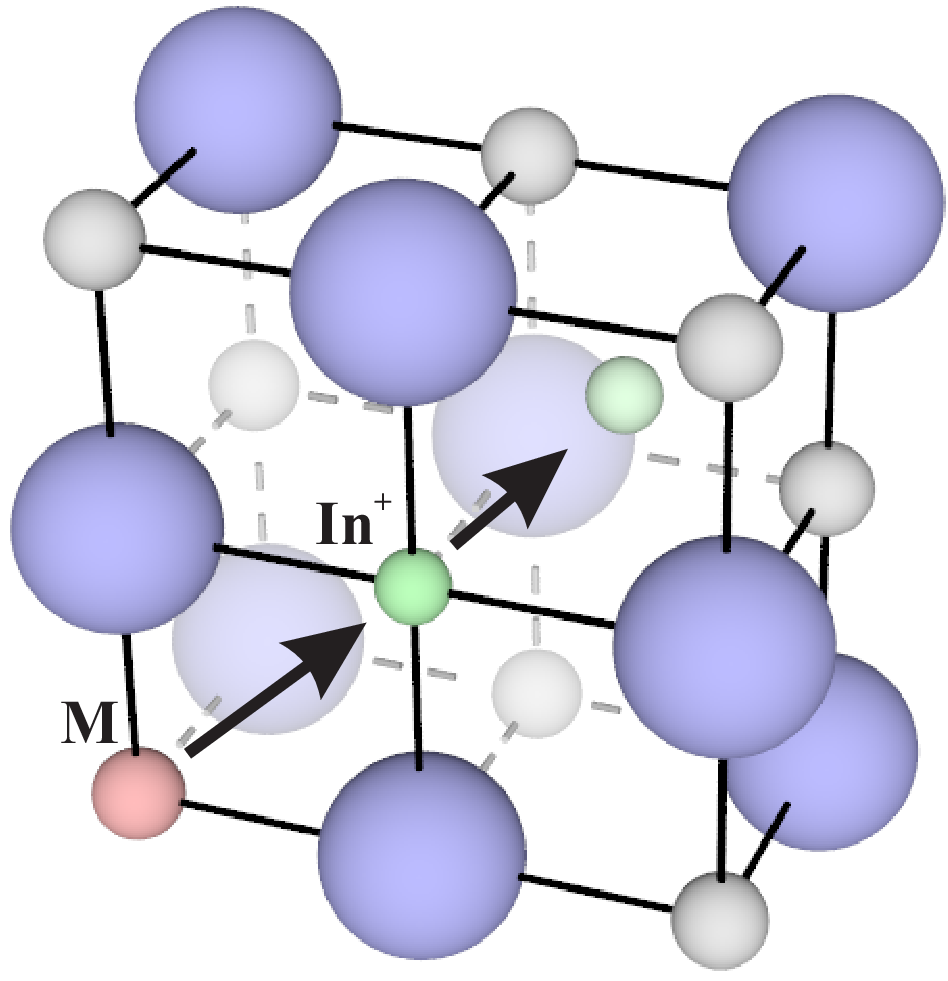}
\caption{In$^+$ cations, which are small in size and carry a low charge, spontaneously form Frenkel defects. 
This involves cations moving to interstitial spaces and leaving behind vacancies.
The formation of Frenkel defects induces the other cations M (= Ag, In, Sn, Pb, Bi) diffusion.}
\label{fig1}
\end{figure}

Focusing on the PbTe-based system, where cations on the Pb site are substituted with Ag, Sn, Bi, and In, we explored ionic crystals resembling the NaCl structure under ambient pressure~\cite{Mizuguchi2023}. The simplicity and clarity of the Coulomb potential in these systems facilitate an in-depth investigation of diffusion over extended periods and large system sizes. Our primary simulations centered on the AgInSnPbBiTe${}_5$ compound. We discovered that In$^+$ cations, which are small in size and carry a low charge, spontaneously form Frenkel defects~\cite{Frenkel1926} (see Fig.~\ref{fig1}). This involves cations moving to interstitial spaces and leaving behind vacancies, significantly impacting diffusion. The study revealed that the charge plays a more crucial role than the size of the cation in the formation of Frenkel defects in the AgInSnPbBiTe${}_5$ system. Furthermore, we observed enhanced diffusion of larger cations near these defects, whereas diffusion was not significantly affected by Schottky defects, where one pair of Pb$^{2+}$ and Te$^{2-}$ is removed. Here we reveal the enhancement of diffusion of atoms in high-entropy compounds, offering insights into the physical properties of HEAs and guiding the development of materials with improved Li-ion permeability.

\section*{Result}
\subsection*{Formation of Frenkel defects}
\begin{figure}[htbp]
\centering
\includegraphics[width=16cm]{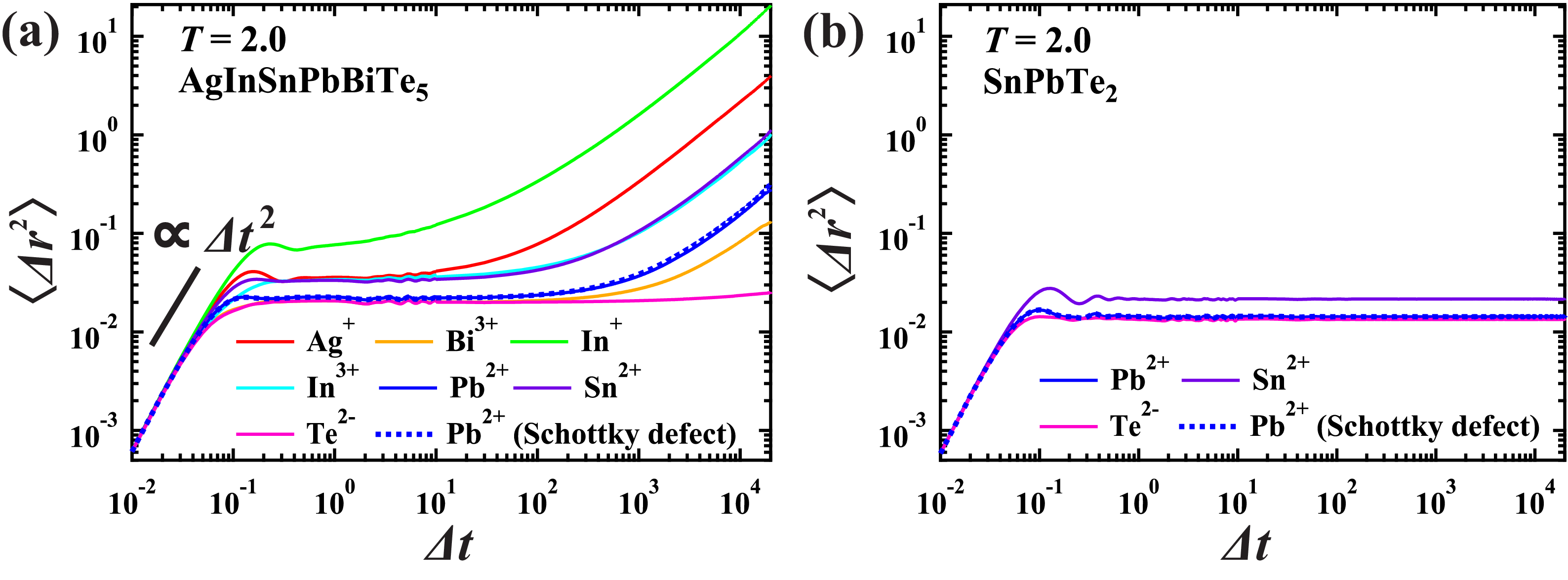}
\caption{$\langle \Delta r^2 \rangle$ in (a) AgInSnPbBiTe${}_5$, (b) SnPbTe${}_2$. 
Following the ballistic regime ($\langle \Delta r^2 \rangle \propto \Delta t^2$), a plateau regime emerges. 
The $\langle \Delta r^2 \rangle$ values during the plateau correspond to intralattice vibration. 
Over extended time periods, the $\langle \Delta r^2 \rangle$ in SnPbTe${}_2$ maintains a plateau, whereas in AgInSnPbBiTe${}_5$, the $\langle \Delta r^2 \rangle$ of cations increases with $\langle \Delta r^2 \rangle \propto \Delta t^\alpha$ ($\alpha \sim 0.9$). 
The dotted line represents the $\langle \Delta r^2 \rangle$ of Pb$^{2+}$ in a system with Schottky defects, where one pair of Pb$^{2+}$ and Te$^{2-}$ is eliminated. 
It is evident that diffusion primarily occurs due to Frenkel defects rather than Schottky defects in AgInSnPbBiTe${}_5$. }
\label{MSD}
\end{figure}

Initially, we conducted an analysis of the time-dependent mean square displacement (MSD) $\langle \Delta r^2 \rangle$ for individual cations to investigate diffusion behavior.
Figures~\ref{MSD}(a) and (b) show $\langle \Delta r^2 \rangle$ of both cations and anions in AgInSnPbBiTe${}_5$ and SnPbTe${}_2$ at $T$ = 2.0, respectively. 
In the short term, $\langle \Delta r^2 \rangle \propto \Delta t^2$ is proportional to $\Delta t^2$, indicating ballistic motion, followed by a plateau observed around $\Delta t \sim 0.1$. 
This plateau corresponds to the magnitude of intralattice vibration, known as the Debye-Wallar factor. 
Notably, $\langle \Delta r^2 \rangle$ for In$^+$ and Ag$^+$ exhibits significant values at the plateau, suggesting larger free space for these cations, while $\langle \Delta r^2 \rangle$ for Bi$^{3+}$ and Te$^{2-}$ indicates tighter packing of these ions. 
Over longer time scales, while $\langle \Delta r^2 \rangle$ of SnPbTe${}_2$ remains constant, that of cations in AgInSnPbBiTe${}_5$ increases in proportion to $\Delta t^\alpha$ ($\alpha \sim 0.9$), indicating diffuse cation movement. 
It is noteworthy that the initial simulation condition assumes a crystal without any atomic deficiencies (Schottky defect), hence the observed diffusion in AgInSnPbBiTe${}_5$ appears exceptional.

\begin{figure}[htbp]
\centering
\includegraphics[width=16cm]{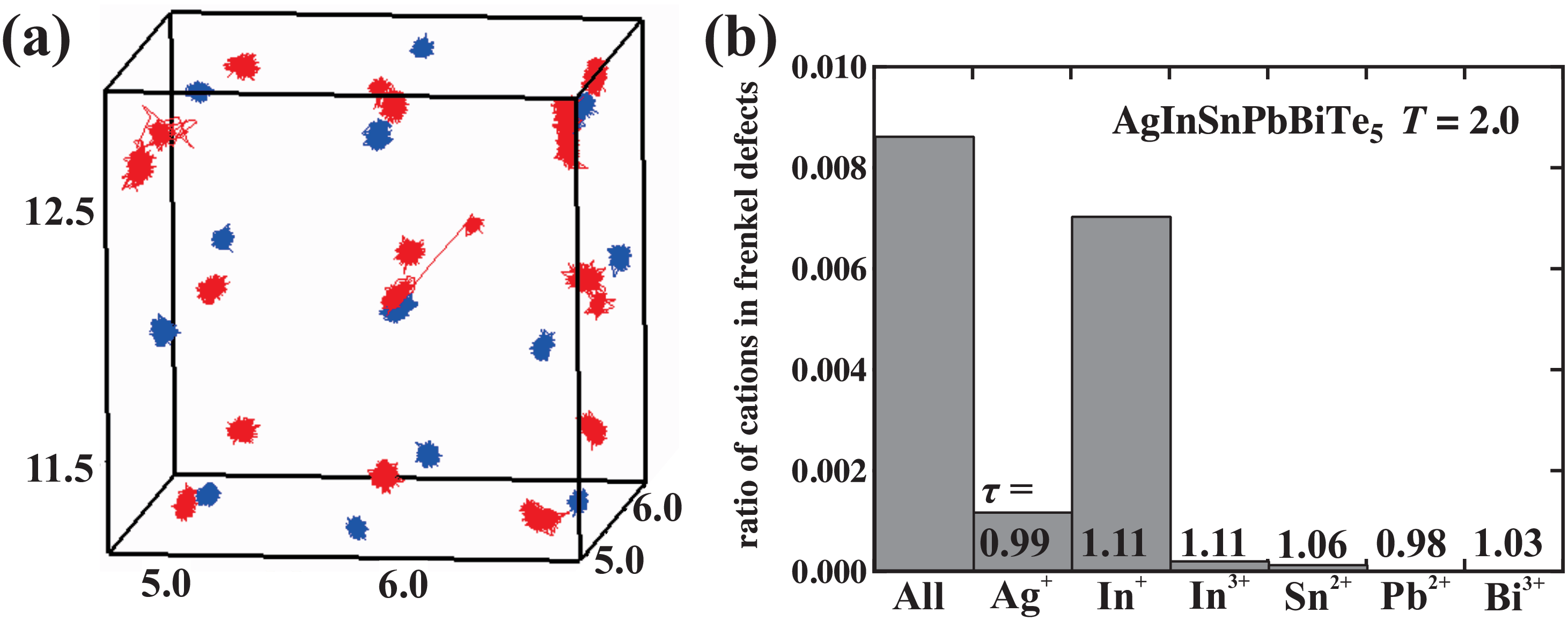}
\caption{(a) Illustrates the trajectories within a single unit cell of AgInSnPbBiTe${}_5$ for $\Delta t$ = 100. 
The trajectories of cations and anions are denoted by red and blue lines, respectively. Notably, the cation positioned at the center transitions to an interstitial location, thereby forming a Frenkel defect~\cite{Frenkel1926}. 
(b) The probability of Frenkel defect formation relative to the total number of cations for each cation type. Our analysis reveals that Frenkel defects predominantly originate from In$^{+}$, followed by Ag$^{+}$. The parameter $\tau$ corresponds to the ratio of free volume in each cation and is defined as $\langle D \rangle /  (\sigma_{\rm{Te}} + \sigma_{\rm{M}})$. 
Given the small $\tau$ value for Ag$^+$, it is evident that the charge plays a more dominant role than cation size in the formation of Frenkel defects.
}
\label{defect}
\end{figure}

In order to elucidate the mechanism of atomic diffusion in these crystals, we conducted an investigation into the trajectories of the particles. Figure \ref{defect}(a) illustrates the particle trajectories within a unit cell over $\Delta t$ = 100 in AgInSnPbBiTe${}_5$ at $T$ = 0.2. 
The trajectories of cations and anions are depicted by red and blue lines, respectively. Our analysis revealed that the cation positioned at the center (In$^+$) transitions to an interstitial position distinct from the stable configuration of the NaCl structure, indicating the formation of a Frenkel defect~\cite{Frenkel1926}. 
Additionally, we investigated the likelihood of Frenkel defect formation at each cation. 
The annealing of AgInSnPbBiTe${}_5$ was conducted at $T$ = 2.0, followed by quenching it at $T$ = 0.1 to observe its inherent structure. 
Figure~\ref{defect} (b) shows the ratio of Frenkel defects to total cations in AgInSnPbBiTe${}_5$. 
Our findings indicate that Frenkel defects predominantly originate from In$^{+}$, followed by Ag$^{+}$.

\subsection*{Local crystal structure}
\begin{figure}[htbp]
\centering
\includegraphics[width=8cm]{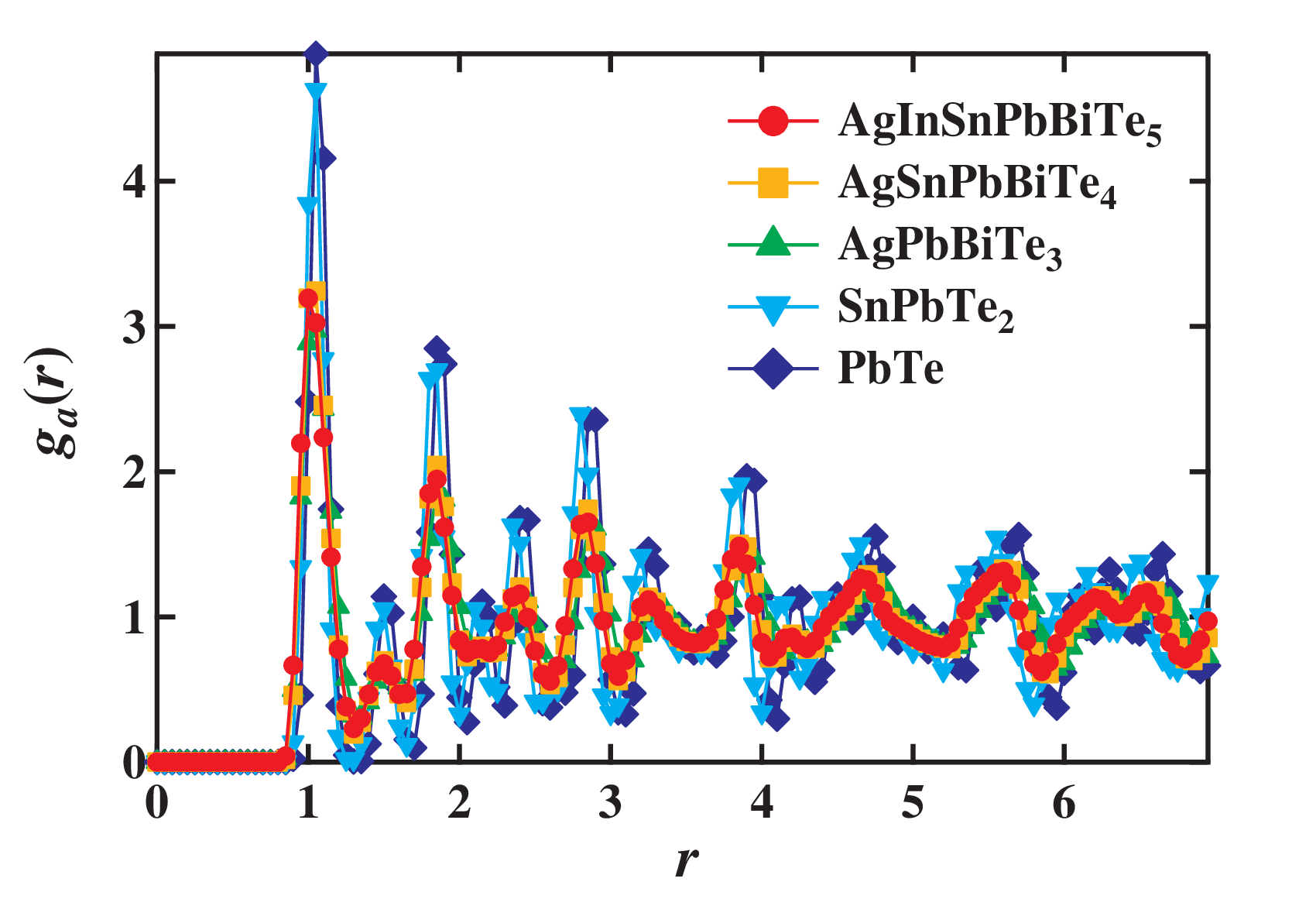}
\caption{Anion-anion radial distribution function $g_a(r)$ at $T$ = 2.0. 
Each symbol represents PbTe, SnPbTe${}_2$, AgPbBiTe${}_3$, AgSnPbBiTe${}_4$, and AgInSnPbBiTe${}_5$, respectively. Notably, the lattice constant in AgInSnPbBiTe${}_5$ closely resembles that of PbTe, with minimal deviation observed. These findings suggest an absence of local distortion in the crystal structure.
}
\label{gr}
\end{figure}

Subsequently, our investigation delved into the structural aspects crucial for Frenkel defect formation. 
To examine the local crystal structure, we computed the anion-anion radial distribution function $g_a(r)$ for crystals in equilibrium. 
Figure~\ref{gr} shows the $g_a(r)$ at $T$ = 2.0 of PbTe, SnPbTe${}_2$, AgPbBiTe${}_3$, AgSnPbBiTe${}_4$, and AgInSnPbBiTe${}_5$. 
Notably, the lattice constant in AgInSnPbBiTe${}_5$ closely mirrors that of PbTe, indicating minimal deviation. These findings align with experimental results~\cite{Mizuguchi2023}, suggesting an absence of local distortion in the crystal structure to mitigate elastic energy loss. 
Furthermore, to assess the available space for each cation, we introduced $\tau = \langle D \rangle / (\sigma_{\rm{Te}} + \sigma_{\rm{M}})$, where $\langle D \rangle$ is the mean lattice constant, $\sigma_{\rm{Te}}$ is a diameter of Te$^{2-}$, and $\sigma_{\rm{M}}$ is a diameter of a cation M (M is Ag, In, Sn, Pb, or Bi). 
As depicted in Figure \ref{defect}(b), $\tau$ values for Ag, In, Sn, Pb, and Bi are 0.99, 1.11, 1.06, 0.98, and 1.03, respectively, indicating substantial free volume. Notably, In$^{+}$, being the smallest in both size and charge, exhibits a propensity for Frenkel defect formation. Interestingly, despite Ag$^+$ being the second largest atom, it also displays a tendency to form Frenkel defects, emphasizing the dominant role of charge in Frenkel defect formation.

\subsection*{Collective movements among cations}
In AgInSnPbBiTe$_5$, $\langle \Delta r^2 \rangle$ of Pb$^{2+}$ exhibits an increase for $\Delta t > 10$, while it remains constant in SnPbTe$_2$.
It’s noteworthy that Schottky defects (atomic deficiencies) are commonly present in crystals, often considered as sites for diffusion. 
Consequently, we evaluated diffusion in a system featuring Schottky defects, where one pair of Pb$^{2+}$ and Te$^{2-}$ is removed. The plotted dashed line in Fig.~\ref{MSD} represents the $\langle \Delta r^2 \rangle$ of Pb$^{2+}$ under these conditions. 
Interestingly, in AgInSnPbBiTe$_5$, the $\langle \Delta r^2 \rangle$ remains unaltered upon introducing the Schottky defect, while the plateau of $\langle \Delta r^2 \rangle$ in SnPbTe$_2$ remains unchanged as well. 
This observation implies a substantially lower hopping probability associated with Schottky defects compared to Frenkel defects.

\begin{figure}[htbp]
\centering
\includegraphics[width=16cm]{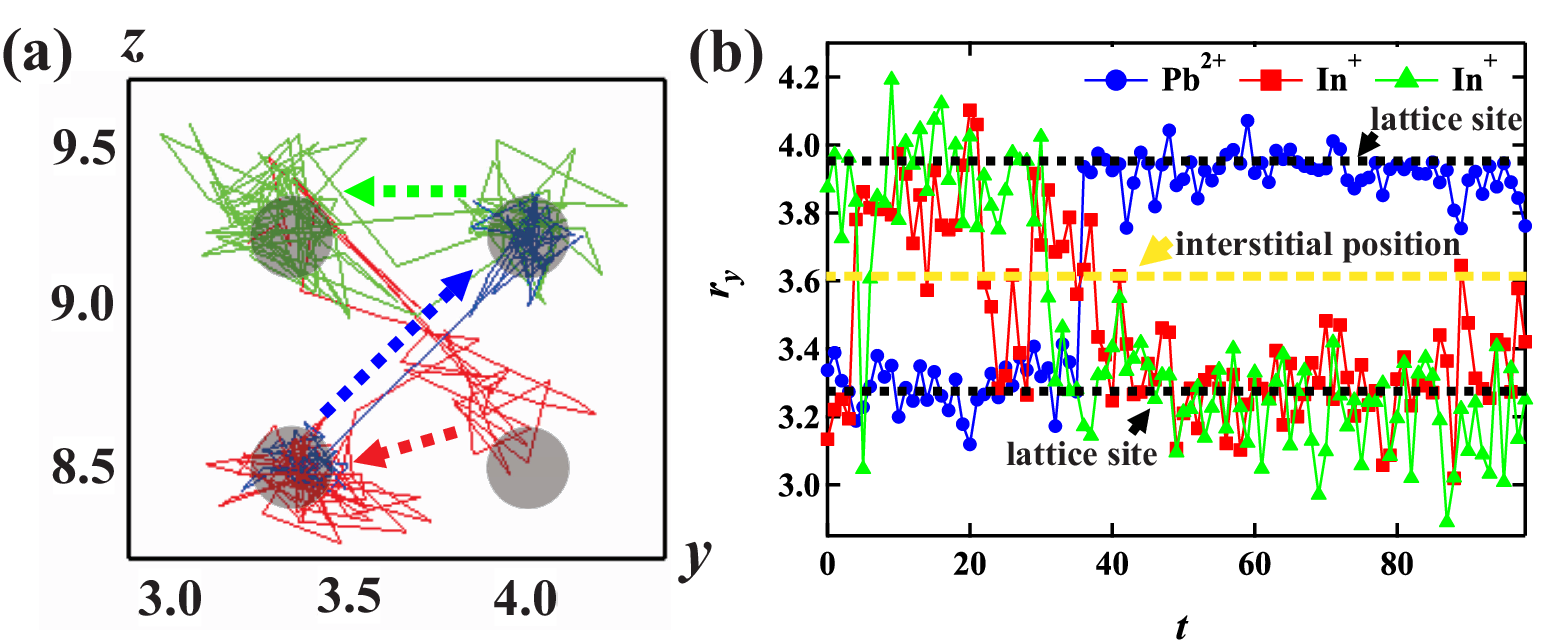}
\caption{(a) Trajectories projected onto the $yz$ surface of the three nearest neighbor cations during $\Delta t$ = 100. 
The trajectories of In$^+$, Pb$^{2+}$, and another In$^+$ are depicted by the red, blue, and green lines, respectively. Dotted arrows serve as visual aids illustrating displacements from the initial position to the position after $\Delta t$ = 100. 
(b) Temporal change of the $y$-coordinates of the three cations. 
It is observed that the two In$^+$ cations transition to other sites concurrently with the hopping of Pb$^{2+}$. This observation suggests interdependent movements among cations, leading to a coordinated diffusion process.
 }
\label{coop}
\end{figure}

The hopping to Schottky defects typically occurs as a single event, as the local energy potential undergoes minimal change during such transitions. Conversely, significant local energy potential changes accompany the formation of Frenkel defects from an energetically stable state. Subsequently, we investigated the local motions of cations surrounding Frenkel defects. Trajectories projected onto the $yz$ surface of the three nearest neighbor cations during $\Delta t$ = 100 are depicted in Fig.~\ref{coop}(a). 
Here, the red, blue, and green lines denote the trajectories of In$^+$, Pb$^{2+}$, and another In$^+$, respectively. It's observed that two In$^+$ cations shuttle back and forth between the Frenkel defect and the lattice site before Pb$^{2+}$ hops to the vacant position. 
Figure \ref{coop}(b) illustrates the temporal evolution of the time evolution of the $y$-coordinates of the three cations, revealing simultaneous movements of the two In$^+$ when Pb$^{2+}$ undergoes hopping. These findings suggest cooperative cation movements influencing each other, leading to diffusion even among cations with larger size and charge.

\subsection*{Formation of short range order}
\begin{figure}[htbp]
\centering
\includegraphics[width=16cm]{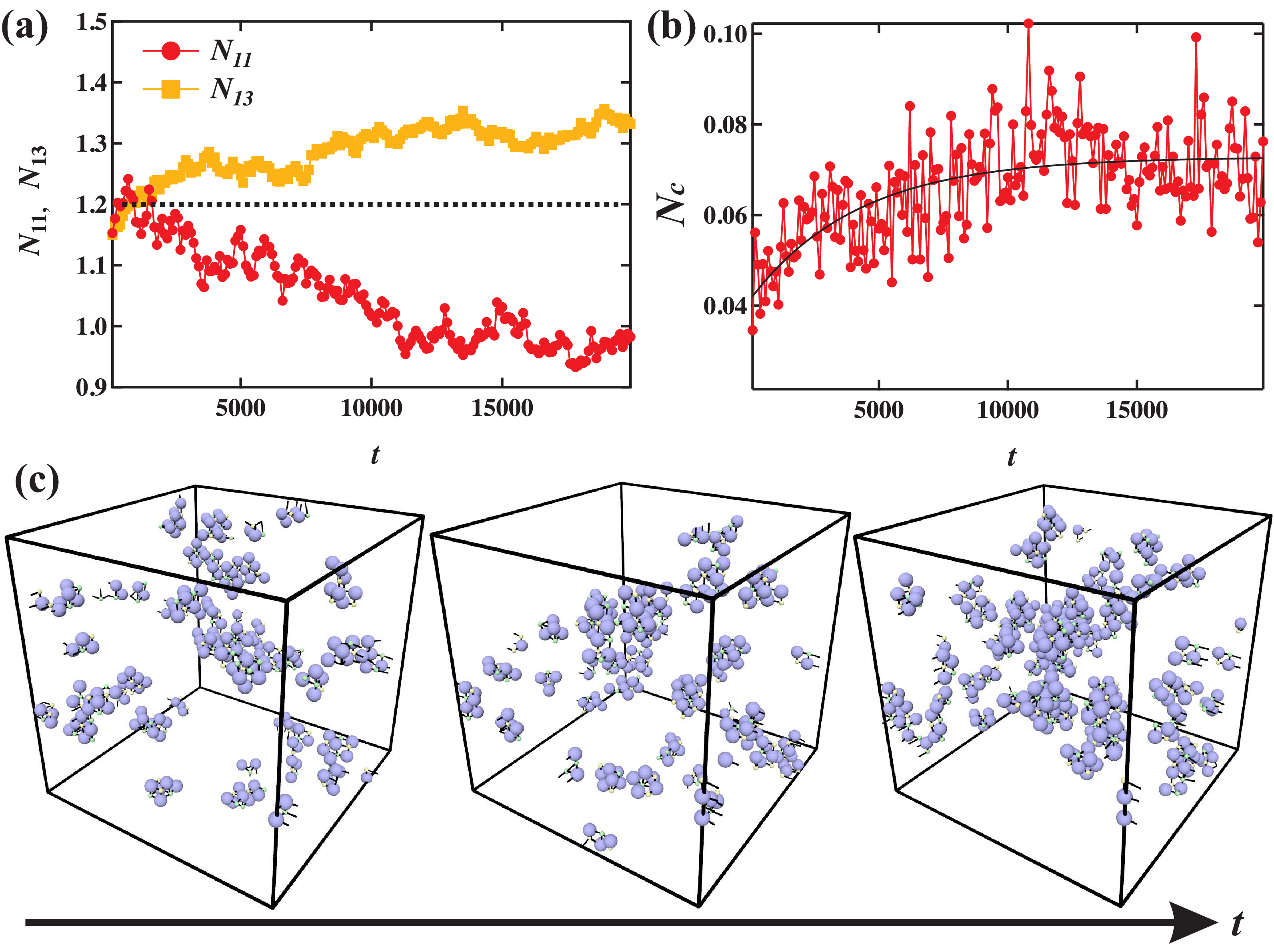}
\caption{(a) 
The temporal changes in the counts of In$^+$, denoted as $N_{11}$, and In$^{3+}$, denoted as $N_{13}$, within the vicinity of In$^+$ are examined. A dotted line represents the anticipated value under random configuration. 
Formation of In$^+$-In$^{3+}$ pairs occurs  to preserve local charge balance. 
In panel (b), the progression over time of the ratio of Te$^{2-}$ within short range order (SRO), denoted as $N_c$, is depicted. SRO is defined as a cluster where the number of loops formed by an In$^+$-In$^{3+}$ pair and Te$^{2-}$ exceeds 5. An exponential fit (black line) yields $N_c = 0.073 -0.032 \exp(-t / 4155.6)$. It is observed that $N_c$ increases over time, indicative of SRO formation. In panel (c), the time evolution of SRO at $t$ = 100, 5000, and 10000 is illustrated. The size of SRO expands with time, reflecting not only an increase in the number of SRO but also their enlargement.
}
\label{NN}
\end{figure}

As cation diffusion becomes more pronounced, the structure undergoes relaxation towards a more stable conformation. In this study, we examined the changes in configuration during the equilibration process in AgInSnPbBiTe$_5$. 
We determined the quantities of In$^+$, denoted as $N_{11}$, and In$^{3+}$, denoted as $N_{13}$, within the vicinity of In$^+$, represented by circle and square symbols in Fig.~\ref{NN}(a). 
Initially, with a random configuration, $N_{11}$ and $N_{13}$ at $t$ = 0 are expected to be 1.2, calculated as the product of the number of nearest neighbor cations and the number fraction of In$^+$, indicated by the dotted line. 
Subsequently, $N_{11}$ decreases over time, while $N_{13}$ increases, signifying the formation of In$^+$-In$^{3+}$ pairs to preserve local charge balance. 
We define short range order (SRO) as a cluster where the number of loops created by an In$^+$-In$^{3+}$ pair and Te$^{2-}$ exceeds 5. 
Figure\ref{NN}(b) depicts the time evolution of the ratio of Te$^{2-}$ within SRO, denoted as $N_c$. An exponential fit (black line) yields $N_c = 0.073 -0.032 \exp(-t / 4155.6)$. Around $\Delta t \sim 4000$, $\langle \Delta r^2 \rangle$ of In$^+$ is substantial (refer to Fig.\ref{MSD}(a)), prompting significant diffusion of In$^+$ cations and subsequent configuration alterations. We also examined the size evolution of the SRO. Figure\ref{NN}(c) illustrates the SRO at $t = 100$, $5000$, and $10000$. 
Notably, the SRO expands over time, indicating not only an increase in the number of SRO but also their enlargement. 
The prolonged growth of SRO may be linked to the aging of High-Entropy Alloys (HEAs), a subject warranting further investigation through extended simulation durations and larger simulation box sizes in future studies.

\section*{Discussion}
First, it is important to note that the ionic diameter of In$^+$ is not well-documented, leading us to assume that In$^+$’s size is comparable to that of In$^{3+}$ despite the expectation that In$^+$ would be larger. To assess the impact of In$^+$’s size on our findings, we conducted simulations with varied sizes of In$^+$, confirming that our results remain essentially unaffected. This supports the notion that charge plays a pivotal role in diffusion processes over the size of the ion. Based on these findings, we propose potential applications of this understanding in designing materials for lithium-ion batteries. Our results indicate that the diffusion mechanism induced by the formation of Frenkel defects would be similar if In$^+$ were replaced with Li$^+$. This suggests that HEA materials incorporating Li$^+$ could exhibit high melting temperatures, significant Li$^+$ diffusivity, and enhanced mechanical strength by fine-tuning the mix ratio and the number of components, making them ideal for use as separators in lithium-ion batteries or all-solid-state batteries to improve safety.

Furthermore, we explore the potential universality of the diffusion enhancement effect observed in multicomponent alloys. Frenkel defects, characterized by a metastable interstitial position within the crystal structure, are not exclusive to NaCl-type ionic crystals but are also present in FCC and BCC structures~\cite{Zhao2019}. The formation and presence of Frenkel defects are crucial in influencing various physical and chemical properties, such as chemical reactions~\cite{Xu2022, Heleen2019}, superionic conductivity~\cite{Keen2003}, optical emissions\cite{Lushchik2000}, atomic trapping\cite{Heleen2019}, and irradiation resistance~\cite{Zhao2019}, emphasizing the importance of controlling these defects in material design. In this study, we demonstrated that In$^+$ spontaneously forms Frenkel defects, suggesting that atoms with smaller sizes and weaker interactions than other atoms in the compound are more prone to creating these defects. This insight reveals that multicomponent crystals not only benefit from easy mixing due to their high configurational entropy but also from an enhanced tendency to form Frenkel defects due to internal bonding inhomogeneities.

Although empirically recognized, the distinct physical properties observed in In-doped materials have lacked a clear mechanism until now. For instance, while PbTe is a semiconductor with excellent thermoelectric properties, the inclusion of In transforms it into a metallic state~\cite{Sawahara2023}. Our simulations indicate that the doping of In leads to the formation of Frenkel defects, which aligns with the general understanding that such defects equate to hole doping, thereby increasing electrical conductivity. This correlation between our simulation results and experimental observations sheds light on the physical properties of In-doped systems and highlights the importance of considering  Frenkel defect formation in understanding these materials' characteristics.

\section*{Summary}
This study has delved into the diffusion mechanisms within high-entropy compounds, focusing on a PbTe-based high-entropy alloy, AgInSnPbBiTe$_5$. Our findings underscore the spontaneous formation of Frenkel defects, where cations migrate to interstitial spaces, leaving vacancies that significantly enhance diffusion. A key discovery is that the charge of the ions plays a more pivotal role in the formation of Frenkel defects than their size. Despite the formation of these defects, the overall lattice structure of AgInSnPbBiTe$_5$ remains consistent with that of PbTe, indicating that the integrity of the crystal lattice is maintained even with the high mobility of In$^+$ ions. This phenomenon is not observed with atomic deficiencies (Schottky defects), which do not contribute to diffusion, highlighting the unique role of Frenkel defects in the atomic dynamics of these materials.

The study further reveals that the diffusion process is accompanied by a relaxation of the crystal structure towards a more stable configuration facilitated by the enhanced mobility of cations. Notably, clusters consisting of In$^+$ and In$^{3+}$ pairs grow over time, indicating the formation of short-range order within the material. This aspect is crucial for understanding the aging process of HEAs and their impact on their mechanical and electrical properties.

Importantly, our results suggest that these findings are not limited to the inclusion of indium; replacing In$^+$ with Li$^+$ would likely yield similar enhancements in diffusion due to the dominant role of charge. This implies that HEA materials containing Li$^+$ could be designed to possess high ion permeability, a high melting temperature, and increased mechanical strength, making them ideal for applications such as separators in lithium-ion batteries and all-solid-state batteries.

Furthermore, our research highlights the widespread presence and significance of Frenkel defects across various crystal structures. These defects are intimately linked to various material properties, including atomic trapping, optical emission, and electrical conductivity. By understanding the spontaneous formation of Frenkel defects in HEAs, we can gain deeper insights into the physical properties of these materials and pave the way for the development of innovative materials with tailored functionalities, such as self-healing capabilities and enhanced thermoelectric performance.

\section*{Methods}
This investigation employed molecular dynamics (MD) simulations to elucidate the mechanisms underlying diffusion and the development of short range order. Particles $i$ and $j$ were subjected to interaction potentials including the Weeks-Chandler-Andersen (WCA) potential $U_{WCA}$ and the Coulomb potentials $U_q$.
\begin{eqnarray}
m \frac{d^2 \vec r_i}{dt^2} &=& \sum_{i\neq j} -\vec \nabla (U_{WCA} (r_{ij}) + U_q (r_{ij})) \\
U_q (r_{ij}) &=& -\frac{k q_i q_j}{r_{ij}}
\end{eqnarray} 
\begin{eqnarray}
U_{WCA} (r_{ij}) &=& \left\{
\begin{array}{ll}
4\epsilon \left[ \left(\frac{\sigma_{ij}}{r_{ij}} \right)^{12} - \left(\frac{\sigma_{ij}}{r_{ij}} \right)^6 \right] +\epsilon & (r_{ij} < 2^{1/6}\sigma_{ij}) \\
0 & (r_{ij} \ge 2^{1/6}\sigma_{ij})
\end{array}
\right.
\end{eqnarray}
where $\vec r_i$, $r_{ij}$, $m$, $\epsilon$, $k$, $q_i$ are the coordinate of particle $i$, the center-to-center distance from particle $i$ to $j$, the mass of particle $i$, the energy coefficient, the Coulomb constant, and a charge of particle $i$, respectively. 
$\sigma_{ij} = (\sigma_i + \sigma_j)/2$, where $\sigma_i$ is a diameter of particle $i$. 
The length scale in our study is normalized using the ionic diameter of Te$^{2-}$ ($\sigma_{\rm{Te}}$ = 4.42\AA\cite{Shannon1976}). Consequently, the values of $\sigma$ for each element, including Ag$^+$, In$^{3+}$, Sn$^{2+}$, Pb$^{2+}$, Bi$^{3+}$, and Te$^{2-}$, are 0.520, 0.362, 0.421, 0.538, 0.466, and 1.00, respectively~\cite{Shannon1976}. It is worth noting that the ionic diameter of In$^+$ is uncertain, so we assume its size to be equivalent to that of In$^{3+}$, although In$^+$ is expected to be larger than In$^{3+}$. We conducted additional simulations by varying the size of In$^+$ and found that the results were essentially unaffected.
In our simulations, the atomic mass $m = 127.6 ~\rm{g/mol}$~\cite{Thomas2022}, and the interatomic interaction strength $\epsilon = 295 k_B \rm{J}$. The value of $\epsilon$ was determined based on the melting point of PbTe (refer to Supplemental Materials). Subsequently, time, temperature, and pressure were measured in units of $t_0 = \sqrt{m\sigma^2_{\rm{Te}}/\epsilon}$, $T_0 = \epsilon/k_B$, and $P_0 = \epsilon/\sigma_{\rm{Te}}^3$, respectively. The unit charge $q_0$ was normalized by $q_0 = \sqrt{\epsilon \sigma} ~ \rm{C}$. For this study, we set $k$ = 1. 
The Coulomb force was computed using the Ewald summation method~\cite{Ewald1921, Frenkel2002}, with the Ewald parameter $\alpha$ = 0.6, a real space cutoff length of 5 $\sigma_{\rm{Te}}$, and a cutoff wavenumber of 7(2$\pi$/L), where $L$ is the system length, ensuring a root mean square error in the force of less than $10^{-4}$.

Our simulation system consisted of $N$ = 13824 particles arranged in a NaCl-type crystal with 12 unit cells in one direction. The cations were randomly distributed in the initial configuration, and the system was equilibrated using the NPT ensemble for an extended period. Subsequently, we transitioned to the NVE ensemble for accuracy, with temperatures remaining constant thereafter. Importantly, the sound velocities and melting points observed in PbTe-based multicomponent systems qualitatively matched experimental results (refer to Supplemental Materials). We conducted more than five simulations starting from different initial configurations, and all results reported in this study were found to be reproducible.

\section*{Acknowledgements}
The authors would like to thank Hiroto Suzuki and Tatsuma D. Matsuda for experimental support.
R. I. was supported by JST SPRING, Grant Number JPMJSP2156. 
K. T. was supported by JSPS KAKENHI Grant Number JP20H05619. 
Y. M. was supported by JSPS KAKENHI Grant Number 21H00151. 
R. K. was supported by JSPS KAKENHI Grant Number 20H01874. 

\section*{AUTHORS CONTRIBUTIONS}
R.~K. and Y.~M conceived the project. R.~I. and K.~T. performed the numerical simulations and analyzed the data. 
R.~I. and R.~K. wrote the manuscript.

\section*{COMPETING INTERESTS STATEMENT}
The authors declare that they have no competing interests. 

\section*{CORRESPONDENCE}
Correspondence and requests for materials should be addressed to R.~I. (ishikawa-rikuya@ed.tmu.ac.jp) and R.~K. (kurita@tmu.ac.jp).

\section*{Availability of Data and Materials}
All data generated or analyzed during this study are included in this published article and its supplementary information files.

%

\pagebreak
\widetext
\begin{center}
\textbf{\large Supplemental Materials for "Spontaneous formation of Frenkel defects in high-entropy-alloys-type metal telluride"}
\end{center}
\setcounter{equation}{0}
\setcounter{figure}{0}
\setcounter{table}{0}
\setcounter{page}{1}
\makeatletter
\renewcommand{\theequation}{S\arabic{equation}}
\renewcommand{\thefigure}{S\arabic{figure}}
\renewcommand{\bibnumfmt}[1]{[S#1]}
\renewcommand{\citenumfont}[1]{S#1}

\section*{Comparison with experiment}
\begin{figure}[htbp]
\centering
\includegraphics[width=8cm]{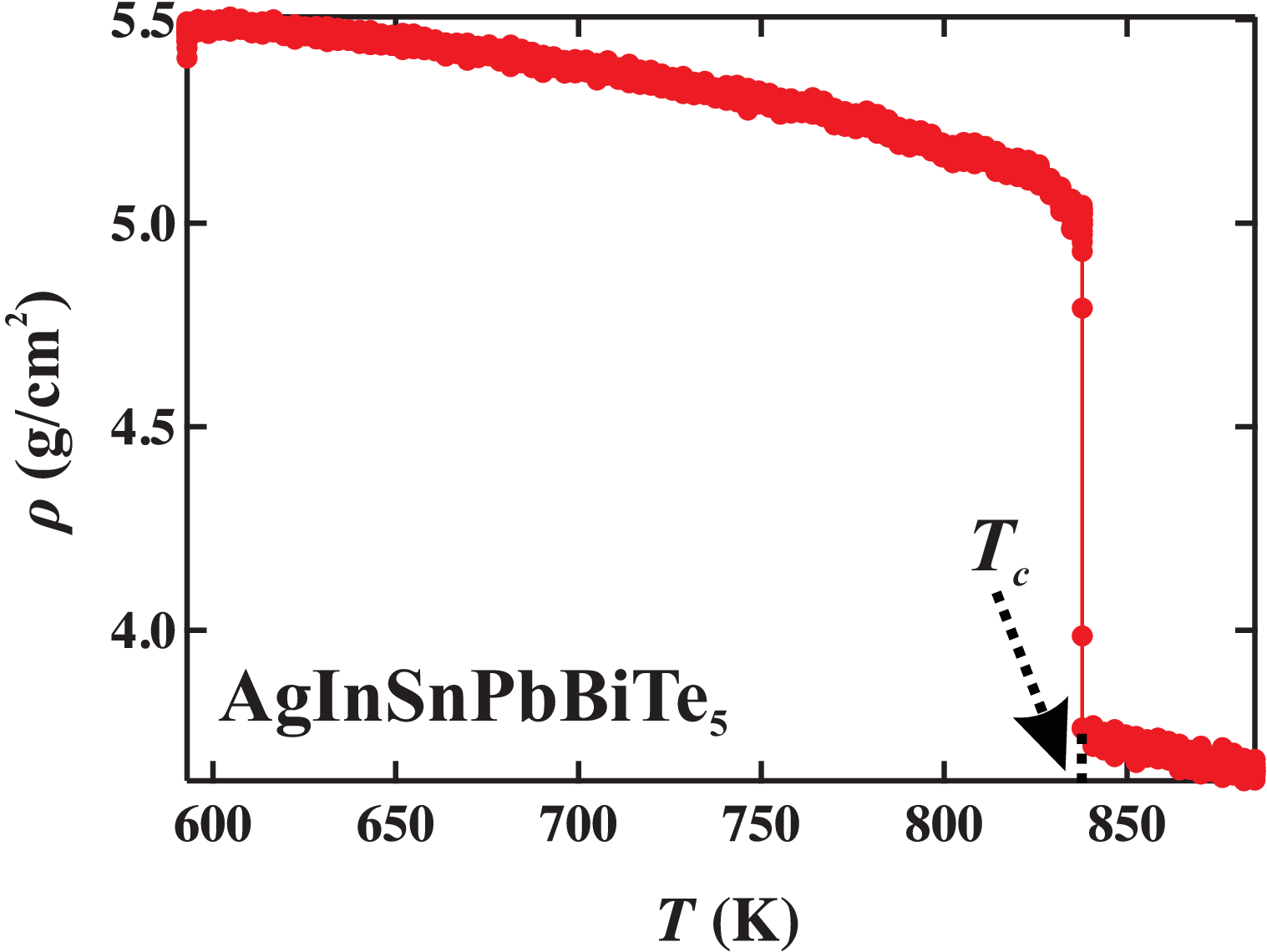}
\caption{(a) Temperature dependence of the density in AgInSnPbBiTe${}_5$. Temperature was increased by 0.01 for every $t$ = 200 in the NPT ensemble. 
The density changes discontinuously at $T_c$. }
\label{melting_point}
\end{figure}

We investigated the melting temperature $T_c$ to compare the numerical simulation with the experiments. 
The temperature was increased by 0.01 for every $t$ = 200 in the NPT ensemble. 
Figure~\ref{melting_point} shows the temperature dependence of the density in AgInSnPbBiTe${}_5$. 
The density decreases with increasing temperature and changes discontinuously at $T_c$. 
We confirmed that the phase transition from crystal to liquid was observed at $T_c$. 
Table~\ref{table_melting_point} shows $T_c$ of PbTe, SnPbTe${}_2$, AgPbBiTe${}_3$, AgSnPbBiTe${}_4$, AgInSnPbBiTe${}_5$ in both simulations and experiments. 
The melting temperatures obtained by simulations and their errors are the mean and standard deviation of five results with different initial conditions.
For AgInSnPbBiTe$_5$, its melting point was obtained experimentally by DSC (NETZSCH, Japan). 
It is confirmed that the simulation results are quantitatively consistent with the experiments. 

\begin{figure}[htbp]
\centering
\includegraphics[width=8cm]{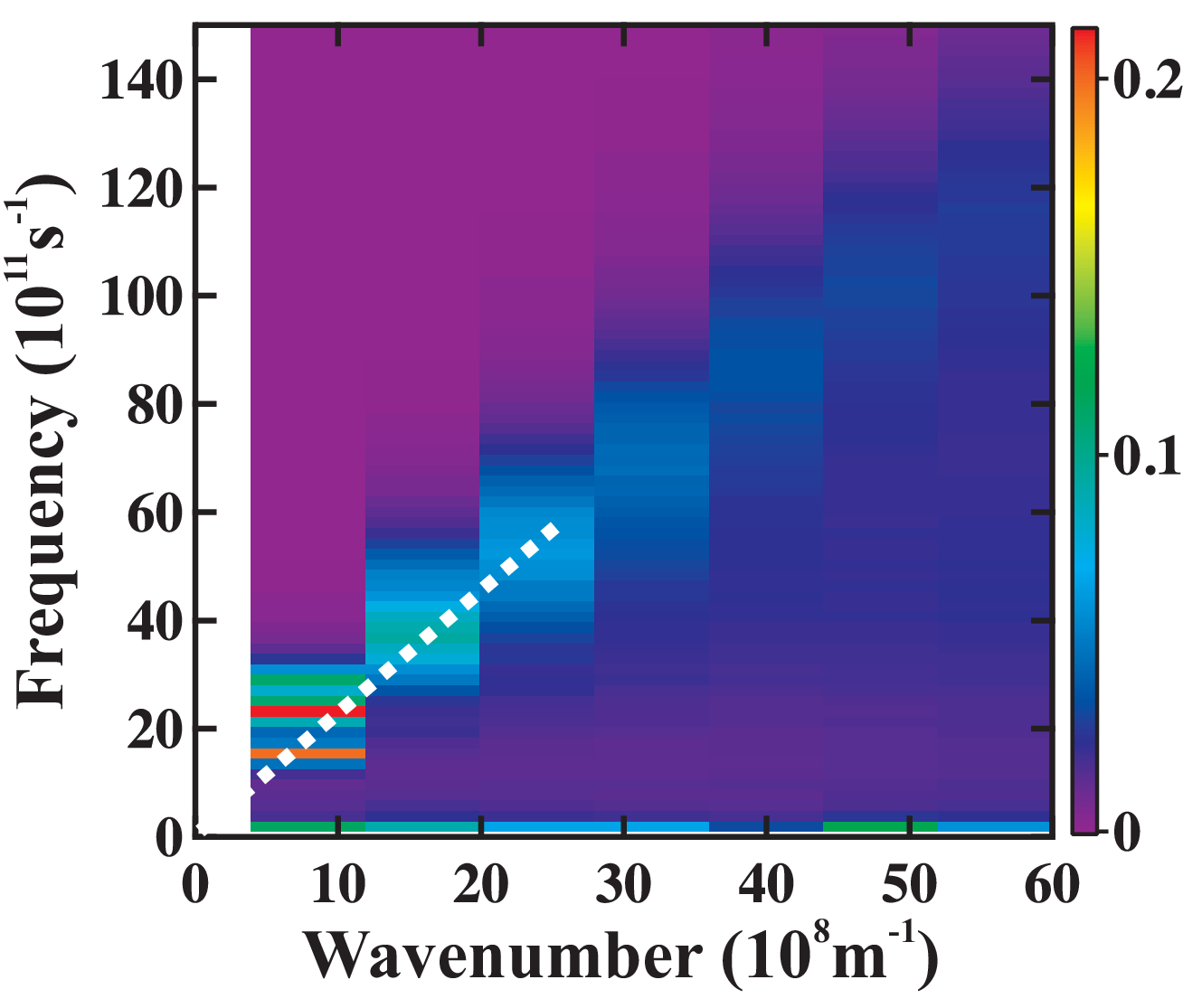}
\caption{Dynamic structure factor of the longitudinal component $S_{vl}$ in AgInSnPbBiTe${}_5$ at $T$ = 1.0. A dispersion relation of acoustic modes is observed on the low wavenumber side. From this slope (white dotted line), the sound velocity of the longitudinal wave was calculated.}
\label{DSF}
\end{figure}

In addition, we calculated the dynamic structure factor of the longitudinal wave and the sound velocity of the longitudinal wave from the dispersion relation of the acoustic modes. 
\begin{eqnarray}
 S_l(k, \omega)=\frac{k^2}{N\omega^2}\int^{\infty}_0dt e^{i\omega t}\langle \vec v_\| (\vec k, t)\cdot \vec v_\| (\vec k, 0)\rangle,
\end{eqnarray} 
where $\vec v_\| (\vec k, t) =(\vec k \vec k /k^2)\cdot \vec v (\vec k, t)$ is the longitudinal component of the particle's velocity and $\vec v (\vec k ,t)=\sum_j\vec v_j(t) \exp(-i\vec k \cdot \vec r_j(t))$. 
Figure~\ref{DSF} shows the dynamic structure factor of the longitudinal component $S_l(k, \omega)$ in AgInSnPbBiTe$_5$ at $T$ = 1.0. 
A liner dispersion relation of the acoustic mode was observed near low wavenumber (white dotted line).

The longitudinal sound velocity was calculated from the dispersion relation of the acoustic modes.
The longitudinal sound velocity at $T$ = 1.0 is shown in Table~\ref{table_melting_point}. 
Since the dispersion relation is broad, the longitudinal sound velocity and its error were defined as the median of the slope and the difference between the maximum and median values. 
The sound velocity also agrees well with the experiment, confirming that the MD simulation reproduces the experimental results well.

\begin{table}[htbp]
 \caption{Melting temperature and the longitudinal sound velocity of each material in simulations and experiments}
 \label{table_melting_point}
 \centering
  \begin{tabular}{c|c|c|c|c|c}
   \hline
  & PbTe& SnPbTe${}_2$ & AgPbBiTe${}_3$ & AgSnPbBiTe${}_4$ & AgInSnPbBiTe${}_5$ \\
   \hline \hline
 \begin{tabular}{c} Melting point (K) \\ (simulation) \end{tabular}  & 1197 & 1138 & 890 & 949 & 834 \\  \hline
  \begin{tabular}{c} Melting point (K) \\(experiment) \end{tabular} & 1197~\cite{David1998} & 1123~\cite{Wagner1967} &  &  &870 \\  \hline
  \begin{tabular}{c} longitudinal sound velocity (m/s)  \\(simulation)\end{tabular} & 2717$\pm148$ & 2525$\pm120$ & 2166$\pm138$ & 2221$\pm165$ & 2139$\pm$96 \\  \hline
   \begin{tabular}{c} longitudinal sound velocity (m/s) \\ (experiment) \end{tabular} & 3596\cite{Pei2012} &  & & & 2740\cite{Abbas2022} \\
   \hline
  \end{tabular}
\end{table}

\end{document}